\documentclass[fleqn,twoside]{article}
\usepackage{espcrc2}
\usepackage{graphicx}

\usepackage{graphicx}
\usepackage[figuresright]{rotating}

\newcommand{\AmS}{{\protect\the\textfont2
  A\kern-.1667em\lower.5ex\hbox{M}\kern-.125emS}}
\def  \abseta {\mid\eta\mid}
\def\ttau{{\tilde\tau}}
\def\lsp{\widetilde\chi_1^0}

\def\tchi{\widetilde\chi}
\def\slashchar#1{\setbox0=\hbox{$#1$}           
   \dimen0=\wd0                                 
   \setbox1=\hbox{/} \dimen1=\wd1               
   \ifdim\dimen0>\dimen1                        
      \rlap{\hbox to \dimen0{\hfil/\hfil}}      
      #1                                        
   \else                                        
      \rlap{\hbox to \dimen1{\hfil$#1$\hfil}}   
      /                                         
   \fi} 
\def\etmiss{\slashchar{E}_T}
\title{Use of Taus in ATLAS}

\author{Ian Hinchliffe\address[MCSD]{ Lawrence Berkeley National
    Laboratory, Berkeley, CA 94720 USA}
        \thanks{This work was supported in part by the Director, Office of Science,
 Office of Basic Energy Research, Division of High Energy
Physics of the U.S. Department of Energy under Contract
DE-AC03-76SF00098}}
       
\begin{document}

\begin{abstract}
At the LHC, new particles can be expected that decay to final states
involving taus. Examples are given from simulations by the ATLAS
experiment showing how such final states can be exploited.
\vspace{1pc}
\end{abstract}

\maketitle

\section{INTRODUCTION}

The total production rate for taus at a hadron collider is not a
useful quantity. Taus must have significant transverse momentum
($p_T$) in order to be observable. Leptonic decays of taus will yield
isolated electrons or muons that can be detected but these can also
be directly produced so discriminating their origin can be
difficult. Hadronic decays of taus result in jets that must be
distinguished from jets arising from QCD processes using the particle
multiplicity and invariant mass. 

\section{PROPERTIES OF TAUS AT LHC}

The dominant standard model production that results in an observable
sample is  $W\to \tau \nu $ which produces  $\sim 1.5 \times
10^8$ events per $10 fb^{-1}$. Given this large sample it is
reasonable to ask if any useful measurements of tau properties can be
made. Measurement of the lifetime is difficult as it requires a
determination of the decay length using the vertex tracking system and
knowledge of the tau momentum. In the case of $Z\to \tau\tau$ the
momentum can be reconstructed using a constrained fit involving the
$Z$ mass (see Section 3)
. The process is statistics limited to a precision of
approximately 1.8 fs for  $30 fb^{-1}$ of data (\cite{TDR} vol I p. 305).  Use of the $W$ decay which has a much larger
rate is more difficult as there are two neutrinos \cite{lifetime}. Two methods were
attempted involving using the observed tau decay products and the tau
mass constraint to determine the momentum of the neutrino in the tau
decay and  an estimator method \cite{Kodama}. In this case the
statistical error could be less due to the larger event
sample provided that the QCD background can be rejected.
 However, the systematic uncertainties from alignment and other
 sources are
difficult to estimate.

Rate decays of the tau can provide a probe of new physics. Lepton
number is known to be violated in the neutrino sector and the rare
decays $\tau \to \mu\gamma$,  $\tau \to 3\mu$ or $\tau\to \mu e^+e^-$
can be expected to occur. In many models 
\cite{Ellis:1999uq} \cite{Feng:2000wt}, \cite{Hisano:1996cp} the first of these  is expected to be
the largest and a simulation will be discussed here \cite{stroy}. 
The signal is an isolated $\mu$ and photon whose
invariant mass reconstructs to the tau. There are two sources 
 of background $\tau\to
\mu\nu\nu\gamma$ and $W\to \gamma\tau\to \gamma\mu\nu\nu$; the latter
dominates. After cuts to reduce the background, the signal acceptance
is approximately $0.5\%$ and the mass resolution is 20 MeV resulting
in a background of approximately 17 events per  $10 fb^{-1}$. The
resulting limit is not competitive with what is possible at Belle and
BaBar unless the full luminosity of the LHC can be exploited. A study
in this environment where the pile-up of minimum bias events degrades
resolution and increases background has not been undertaken.

\section{RECONSTRUCTION OF $Z\to \tau\tau$}

\begin{figure}[htbp]
\includegraphics[width=6cm]{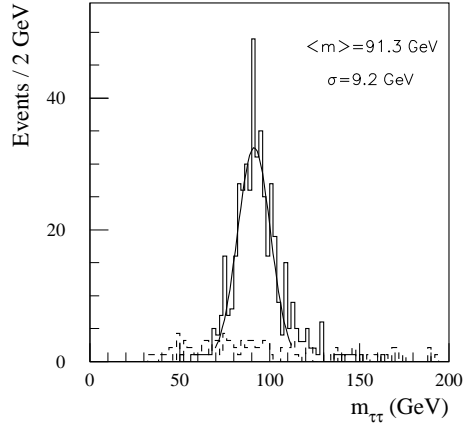}
\caption{Reconstructed mass of the  $Z\to \tau\tau$ signal for $30
  fb^{-1}$. The dashed line shows the background. From \cite{TDR},
    figure 9.54 \label{fig1}}
\end{figure}

The  $H\to \tau\tau$ process is important as a tool for searching for
Higgs bosons at LHC. As a prelude to this and to illustrate the
technique, I will first discuss  $Z\to \tau\tau$. Missing neutrinos
imply that tau energy cannot be measured directly. However
 the direction of tau can be obtained from observed decay products as
 the energy of the produced taus is large compared to the mass. If
 $Z$ has signifificant transverse momentum  so that tau's are not back
 to back
in phi  and the only
missing $P_T$ in the event  arises from tau decay, 
then the tau momenta and invariant mass of the $\tau\tau$ system 
can be reconstructed by assuming that the neutrinos follow the
direction of the observed decay products.
 These events have no other features so that the
selection must be based on the pair of taus. There is a very large
di-jet background from QCD events which must be overcome and the
events require a trigger. Events are selected by requiring an isolated
electron or muon with $P_T>25$ GeV and $\abseta <2.5$  and hadronic
jet 
with $E_T>30 GeV$,  $\abseta <2.5$, and a number of associated
reconstructed tracks  $N_{track}=1 \hbox{
  or }3 $, The jet is also required to be narrow, having its energy deposits in the
electromagnetic calorimeter confined with a cone of radius
$R_{em}=0.07$ in $\eta-\phi$ space.
  Figure \ref{fig1} shows the reconstructed $Z$ mass with the peak at
  the correct value. The figure shows
  the mass resolution of $\sim 10\%$ which is dominated by the missing
  $E_T$ resolution of the detector.  The  small residual QCD
  background is also shown. The subset of events with $N_{track}=3$ can be used to
  determine the tau decay vertex for the lifetime measurement
  discussed above.

\section{NEUTRAL HIGGS SEARCH}

The decay $H\to \tau\tau$ can be reconstructed using the technique
described in the previous section and the mass of the Higgs
measured. The production rates and branching ratios of the standard
model higgs are too small for the signal to be seen above the QCD
background using the dominant production process $gg\to H$.
However the lower rate process $qq\to qqH \to qq\tau\tau$ while it may
not enable a discovery to be made, will give information on the Higgs
couplings. The two jets arising from the quarks in the production
process are at large rapidity and can be used to reject
background. Final states $\tau\tau\to \ell^\pm h^\mp\ell p_T$ and
  $\tau\tau\to \to e^\pm \mu^\mp$ are used and the $\tau\tau$
  invariant mass reconstructed as above with a resolution  $\sigma \sim
 0.1 M$. The table shows the expected signal and
 background \cite{Djouadi:2000gu}. It is important to emphasize that this result is based on
 a parametrized simulation that assumes that the tau tagging and jet
 tagging is not degraded in the presence of pile-up. Nevertheless it
 indicates a
 viable signal for  mass range $110 \hbox{ GeV} <M_H<150\hbox{ GeV}$ 
if the Higgs mass were already known
 from another process. Note that signal to background ratio is such
 that structure would clearly be seen at the Higgs mass.

\begin{table}
\vspace{0.3in}
\caption{ Number of expected signal and background events for the
$qq\to qqH\to\tau\tau jj$ channel, for 100~${\rm fb}^{-1}$ and two
detectors. Cross sections are added for
$\tau\tau\to \ell^\pm h^\mp\ell p_T$ and $\tau\tau \to e^\pm \mu^\mp $
 }
\vspace{0.15in}
\begin{tabular}{c|cccccc}
    $m_H$ &      100  & 110  &  120  &  130  &   140  &  150  \\
\hline
          Signal &   211  &   197  &   169  &   128  &    79  &   38   \\
          Background &   305  &  127  &   51   &    32  &    27  &  24   \\
\end{tabular}
\end{table}

\begin{figure}[htbp]
\includegraphics[width=6cm]{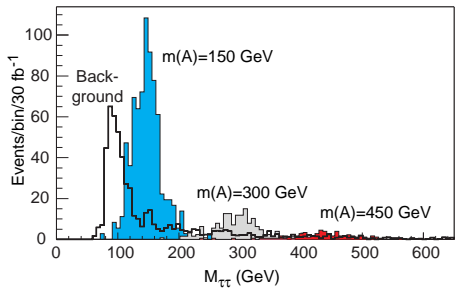}
\caption{Reconstructed mass of the  $A\to \tau\tau$ signal for $30
  fb^{-1}$ showing three possible masses.  The  background is shown as
  the unfilled histogram. From \cite{TDR}. \label{atau}
    }
\end{figure}

In extensions to the standard model more higgs bosons are
expected. The minimal supersymmetric model (MSSM) has three neutral
($h,A,H$) and one charged ($H^\pm$) particles. Their properties are
determined by two parameters which can be taken to be the mass of $A$
and $\tan\beta$, which controls the coupling to tau's. $A$ and $H$ can be
observed over certain regions of parameter space. The technique
adopted is the same as that used for $Z\to\tau\tau$. Figure~\ref{atau}
shows  three possible signals and the background which arises from QCD
jets. This signal covers a large fraction of MSSM parameter space and
is one of the most important probes of the MSSM Higgs sector.

\label{tausec}

\section{CHARGED HIGGS}

\begin{figure}[htbp]
\includegraphics[width=6cm]{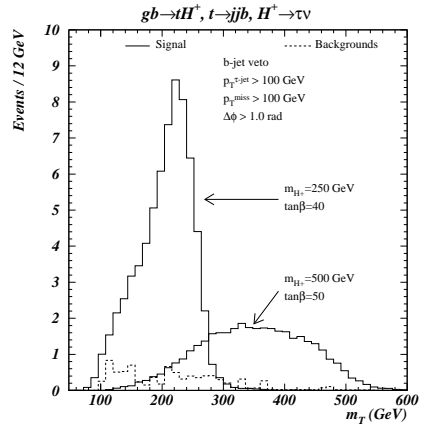}
\caption{Transverse  mass of the Missing $E_T$ and the jet identified
  as a tau.  The signal from $H\to \tau\nu$ shows 
  two possible masses.  The  background is also shown. From \cite{Assamagan:2002ne}. \label{htau}
    }
\end{figure}

A charged Higgs boson is expected to decay $H^+ \to \tau \nu$ or $H^+
\to q \overline{q}$. The latter decay involves the reconstruction of
jets and the QCD background in the di-jet mass distribution is a
serious difficulty to observation of this mode. The former decay is
more promising. Production of a heavy $H^\pm$ via
$q\overline{q}\to H^+H^-$ is too small, the   production via $gb\to
H^-t$ has $\sigma (H^+t)\  \sim 1$ pb for a mass of $H^\pm$ of 400
GeV. The decay  $H^+ \to \tau \nu$ will result in either an isolated
lepton or jet. The former is not useful so the latter is
exploited. The final state $Ht \to \tau\nu W b$ is 
identified by requiring three jets one of which is tagged as
containing a b-hadron from the vertex system and requiring that these
three jets are consistent with $t\to bW\to b q\overline{q}$  ($m(jj)=M_W\pm 25$ and  $m(jjb)=M_t\pm 25$). Another jet with only one
associated charged track and $p_T>100$ GeV is required along with 
missing $E_T$ ($>100$ GeV). The signal is in transverse mass of
``$\tau$'' and missing $E_T$ which is shown in Figure~\ref{htau}. The
peak is displaced below the Higgs mass as the missing $E_T$ has
contributions from both neutrinos in the decay  $H^+ \to \tau \nu \to
\pi \nu\nu $ and these tend to cancel.

\section{TAUS IN SUPERSYMMETRY EVENTS}

If supersymmetry 
(SUSY) is relevant to the problems of electroweak symmetry breaking,
SUSY particles will be sufficiently light that   the LHC will be a SUSY factory.
Third generation sparticles  are particularly interesting as they 
carry information about how supersymmetry is broken.
In the case of the first two generations the mixing between the
supersymmetric partners of the left and right handed charged leptons
is expected to be small. However the 
two $\ttau$ mass eigenstates are mixtures of  $\ttau_L$ and $\ttau_R$ and
we need to measure  masses and mixings.
The direct  production of $\ttau$ via electroweak processes such as
$q\overline{q} \to \tau^+\tau^-$ is small. In many cases many more
$\ttau$'s are produced from decays of  
strongly produced  squarks and gluinos via processes such as 
$\tilde q \to \tchi_2 q \to  q \ttau^+ \tau^- \to q \lsp \tau^+
\tau^-$ where  $\tchi_2$ and $\lsp$ are weak gauginos and the latter
is stable and exits the detector contributing to missing $E_T$. It is
possible that taus may be the only
  leptons  produced in gaugino decay.

 Leptonic tau decays are of limited use as the origin of the lepton is
 unknown; one must search for hadronic tau decays that give rise to
 characteristic jets. SUSY events are expected to be very complex and
 one can rely on jet  and $\etmiss$ cuts to get rid of standard
 model  background and
obtain clean SUSY sample. This sample can then be studied in
detail. The   background to hadronic tau decays then arises from QCD
jets in the SUSY event and a rejection factor of order 10 (the number
of jets in a such an event) rather than order 100 which is needed in
the analysis described in section \ref{tausec}.

\begin{figure}[htbp]
\includegraphics[width=6cm]{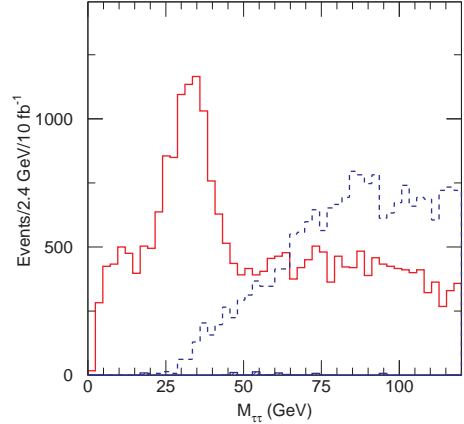}
\caption{Invariant mass of the two tau candidates in SUSY events; the
 solid histogram shows the signal while the dashed shows the  
     background from the  combination of real tau and misidentified
     QCD jet, the small solid shows the  background from QCD jets were
     both are misidentified.
  \cite{TDR}, Figure 20-42. \label{tautau}
    }
\end{figure}

A particular model is used in the  example illustrated here.
 (\cite{TDR} vol II p. and \cite{Hinchliffe:1999zc}).
 Events are selected as follows:
$\ge 4$ jets, one of which  has $p_t>100$ GeV and the rest have
$p_t>50$ GeV, there are no  isolated electrons or muons  with $p_t>10$
GeV and 
$\etmiss>100$ GeV. The jets are examined and ``tau'' candidates
extracted by selecting jets with low track multiplicity and invariant
mass \cite{taufull} using an algorithm based on a study using $Z$
decays to generate taus in the appropriate kinematic regime. 
The invariant mass of observed ``tau'' pairs is then shown in
Figure~\ref{tautau}. A clear peak can be seen above the background
which is dominated by events where one of the tau jet candidates is a
misidentified QCD jet.  If all the tau energy
had been detected, the distribution would have had a sharp cut-off at
$$
M_{\tau\tau}^{\rm max} = M_{\tilde\chi_2^0} 
\sqrt{1-{M_{\tilde\tau_1}^2 \over M_{\tilde\chi_2^0}^2}}
\sqrt{1-{M_{\lsp}^2 \over M_{\tilde\tau_1}^2}} 
$$
which evaluates to  61 GeV in this example. The peak is below this due
to the energy carried of by neutrinos in the tau decays.

The situation is rather more complicated than this analysis shows. The
energy distribution of the visible tau decay products depends on the
tau polarization and the dependence is largest in the $\tau \to
\pi\nu$ final state and smallest in the states with large hadron
multiplicity. A separation of these states could  therefore yield
information on the polarization of the taus. This in turn gives
information on the $\ttau$ composition. Simulation work on this
difficult topic is underway.

\end{document}